# Lumen boundary detection using neutrosophic c-means in IVOCT images

Mohammad habibi
Department of Electrical Engineering
Iran University of Science and Technology
Tehran, Iran
Email: mohammad_habibi@elec.iust.ac.ir

Ahmad Ayatollahi
Department of Electrical Engineering
Iran University of Science and Technology
Tehran, Iran
ayatollahi@iust.ac.ir

Niyoosha Dallalazar
Department of Electrical Engineering
Iran University of Science and Technology
Tehran, Iran
n_dallal@elec.iust.ac.ir

Ali Kermani
Department of Electrical Engineering
Iran University of Science and Technology Tehran, Iran
a_kermani@elec.iust.ac.ir

*Abstract*— In this paper, a novel method for lumen boundary identification is proposed using Neutrosophic c_means. This method clusters pixels of the intravascular optical coherence tomography image into several clusters using indeterminacy and Neutrosophic theory, which aims to detect the boundaries. Intravascular optical coherence tomography images are cross-sectional and high-resolution images which are taken from the coronary arterial wall. Coronary Artery Disease cause a lot of death each year. The first step for diagnosing this kind of diseases is to detect lumen boundary. Employing this approach, we obtained 0.972, 0.019, 0.076 mm$^2$, 0.32 mm, and 0.985 as mean value for Jaccard measure (JACC), the percentage of area difference (PAD), average distance (AD), Hausdorff distance (HD), and dice index (DI), respectively. Based on our results, this method enjoys high accuracy performance.

I. INTRODUCTION

Atherosclerotic cardiovascular disease (CVD) [1], otherwise known as Coronary artery disease [2], is a long-term illness which improves in one's life and usually reaches to a progressed level by demonstrating its' signs [1]. This illness is mainly caused by the integration of atherosclerotic plaques on the coronary artery wall, and as a result of which the lumen area decreases [2]. One evaluation proposed that 80% of all CVD death happens in developing countries [1]. intravascular optical coherence tomography (IVOCT), which is one of the most precise imaging technologies can afford cross-sectional images and micrometer-scale [3]. IVOCT can capture the most casual layers of the arterial wall in addition to the stent struts, and the vascular tissue surrounding them, and lumen region [4].

Nowadays, the majority of studies based on IVOCT images are fulfilled manually. On the other hand, considering large datasets is only possible by employing automatic methods [5]. Automatic Lumen boundary detection methods are divided into three categories, including region-based methods, edge-based methods, and learning based method [6, 7]. Region-based methods including statistical [8] and graph-cut methods [9], edge-based methods such as active contour [4], and learning-based methods including neural networks [10], deep learning [7, 11] and clustering [12, 13].

Data clustering is a significant method in pattern recognition, machine intelligence, and computer science [14]. Data clustering aims to identify identical data such as a set of patterns, points, or objects in a group. Cluster analysis runs without using category labels which define objects by prior identifiers, i.e., class labels [15]. According to most of the literature, the clustering algorithms can be divided into hard and fuzzy clustering methods [16, 17]. Some clustering methods include FCM method and Neutrosophic c_means (NCM) method and so forth [16, 18], And we propose a new clustering method in IVOCT images.

In this paper, an automatic method is proposed for IVOCT images analysis. Employing the image clustering method based on NCM [16], the automatic detection algorithm is used for vessel lumen border identification. The main purpose of this work is to detect boundary as well as outlier points in IVOCT images. To do so, indeterminacy in Neutrosophic (NS) domain which is followed by introducing this set for NCM clustering cost function in IVOCT images is proposed. Using the indeterminacy of data is very useful for boundary detection. However, Methods for lumen boundary detection in most of the papers do not consider indeterminacy. In this paper, we propose a method based on indeterminacy and NS theory for lumen border detection in IVOCT images. The NCM method first applies the NS objective function to the image then apply a mean filter for.

## II. PRELIMINARIES

The method which proposed in this paper is based on NS and Fuzzy clustering. In this section, we are going to introduce these approaches and review them.

### A. FCM

The Fuzzy C Means (FCM) is the first fuzzy method in which the regions of an image are clustered [19]. In this method, the probability that a pixel belongs to a particular cluster is computed by using the membership function [20] which is denoted in (1). This equation represents a cost function that is minimized iteratively in the optimization process to improve the accuracy.

In the abovementioned formula, the membership of pixel $x_j$ in the $i^{th}$ cluster is illustrated using $u_{ij}$, the center of $i^{th}$ the cluster is shown by $v_i$, a norm metric and a constant variable are depicted by $\|\ \|$ and m, respectively. The fuzziness of the resulting partition can be controlled through the variation of the m parameter. Equations (2) and (3) illustrate the updated formulas for the membership functions and cluster centers.

$$u_{ij} = \frac{1}{\sum_{j=1}^{N} \left(\frac{\|X_j - v_i\|}{\|X_j - v_k\|}\right)^{\frac{2}{m-1}}} \quad (2)$$

$$v_i = \frac{\sum_{j=1}^{N} u_{ij}^m X_j}{\sum_{j=1}^{N} u_{ij}^m} \quad (3)$$

Firstly, we guess the cluster center for each of the clusters. The local minimum of the cost function is found by converging the FCM for $v_i$. At each two successive iteration steps, we can assess the differences in the membership function or the cluster center to detect the convergence [20]. As it is shown in [21], the FCM methods are highly influenced by the noise presence in the input data.

### B. Neutrosophic

One of the newly brought up branches of philosophy is called Neutrosophy. This theory contemplates A in a correlation with its opposite, Anti-A, and the neutrality of it, Neut-A, which is neither A nor Anti-A [22]. NS describes a pixel P in the image as $p(t, i, f)$ as follows: t is the true percentage, i is the indeterminacy percentage, and f is the false percentage. The image pixels are turned into the NS domain which is presented by $P_{NS}(i,j) = \{T(i,j), I(i,.,j), F(i,j)$ where t varies in T, i varies in I, and f varies in F. The membership values shown by $T(i,j), I(i,j)$ and $F(i,j)$ are as follows [18, 23]:

$$T(i,j) = \frac{\bar{g}(i,j) - \bar{g}_{\min}}{\bar{g}_{\max} - \bar{g}_{\min}} \quad (4)$$

$$\bar{g}(i,j) = \frac{1}{w \times w} \sum_{m=i-\frac{w}{2}}^{i+\frac{w}{2}} \sum_{n=j-\frac{w}{2}}^{j+\frac{w}{2}} g(m,n) \quad (5)$$

$$I(i,j) = \frac{\delta(i,j) - \delta_{\min}}{\delta_{\max} - \delta_{\min}} \quad (6)$$

$$\delta(i,j) = abs(g(i,j) - \bar{g}(i,j)) \quad (7)$$

$$F(i,j) = 1 - T(i,j) \quad (8)$$

In here, $g(i,j)$ shows the intensity value of the pixel $P(i,j)$, the local mean value of $g(i,j)$ is illustrated by $\bar{g}(i,j)$, and $\delta(i,j)$ is the result of the absolute value of the difference between $g(i,j)$ and $\bar{g}(i,j)$.

## III. METHOD

Based on the NCM method which combines the FCM approach and the NS theory, we propose the lumen border detection methodology which contains two main stages: the preprocessing stage and the employment of the Lumen detection using NCM method in the second stage (Fig.1).

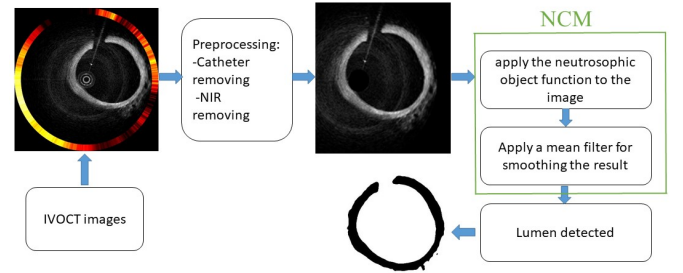

Fig.1. The partitioning of IVOCT lumen identification

### A. Lumen detection using NCM method

In IVOCT image clustering we are looking for boundaries which are indeterminate parameters. The methods that have been used in previous works didn't use indeterminacy for lumen border detection in IVOCT images. We implement a boundary detection algorithm which is able to use indeterminacy. FCM is highly influenced by the noise presence which is solved in NCM by using the NS theory. In this method, at first, we apply the NS objective function to the pixels of an image and iteratively decrease the cost of the function. The three main equations, namely T, I and F, determine the membership degree to the determinant clusters, an ambiguity cluster, and an outlier cluster for each data point, respectively. Secondly, by using a mean filter we can smooth the results. IVOCT images include some noises such as the speckle noise. As the FCM method is highly influenced by the speckle noise, clustering the images will not be performed properly. This can be addressed through the NCM method that aims to detect boundary, employ indeterminacy and eliminate the noise which leads to an accurate and effective clustering. An objective function and membership are defined [16] to consider the indeterminacy in the clustering which leads to the definition of the following equations:

$$L(T,I,F,C) = \sum_{i=1}^{N}\sum_{j=1}^{C}(\bar{\omega}_1 T_{ij})^m \|X_i - C_j\|^2$$
$$+ \sum_{i=1}^{N}(\bar{\omega}_2 I_i)^m \|X_i - \bar{C}_{i_{max}}\|^2 + \sum_{i=1}^{N}\delta^2(\bar{\omega}_3 F_i)^m \quad (9)$$

$$\bar{C}_{i_{max}} = \frac{C_{pi} + C_{qi}}{2} \quad (10)$$

$$P_i = \arg\max(T_{ij}) \quad j = 1,2,\ldots,C \quad (11)$$

$$q_i = \arg\max(T_{ij}) \quad j \neq pi \cap j = 1,2,\ldots,C \quad (12)$$

In equation (9), m has a constant value, and in equations (11) and (12) pi and qi are the clusters number which enjoy the biggest and the second biggest quantity of T, respectively. For each data point i, a constant value $\bar{C}_{i_{max}}$ is calculated after the values for the pi and qi are identified [16]. In here the equations for $T_{ij}$, $I_i$, and $F_i$ are presented:

$$T_{ij} = \frac{K}{\bar{\omega}_1}(X_i - C_j)^{-(\frac{2}{m}-1)} \quad (13)$$

$$I_i = \frac{K}{\bar{\omega}_2}(X_i - \bar{C}_{i_{max}})^{-(\frac{2}{m}-1)} \quad (14)$$

$$F_i = \frac{K}{\bar{\omega}_3}\delta^{-(\frac{2}{m}-1)} \quad (15)$$

$$C_j = \frac{\sum_{i=1}^{N}(\bar{\omega}_1 T_{ij})^m X_i}{\sum_{i=1}^{N}(\bar{\omega}_1 T_{ij})^m} \quad (16)$$

$$K = \left[\frac{1}{\bar{\omega}_1}\sum_{j=1}^{c}(X_i - C_j)^{-(\frac{2}{m}-1)} + \frac{1}{\bar{\omega}_2}(X_i - \bar{C}_{i_{max}})^{-(\frac{2}{m}-1)} + \frac{1}{\bar{\omega}_3}(\delta)^{-(\frac{2}{m}-1)}\right]^{-1} \quad (17)$$

Hence, the clustering is fulfilled by using an iterative optimization method, and in each iteration, the membership $T_{ij}, I_i, F_i$ and the cluster centers $C_j$ are updated. According to the indexes of the largest and the second largest value of $T_{ij}$ at each iteration, the $\bar{C}_{i_{max}}$ is calculated. The iteration continues until $|T_{ij}^{(k+1)} - T_{ij}^{(k)}| < \varepsilon$, k shows the number of iteration steps, and ε is a termination criterion which is a number in the 0 to 1 interval. The following figure illustrates how our framework detects the lumen in an IVOCT image.

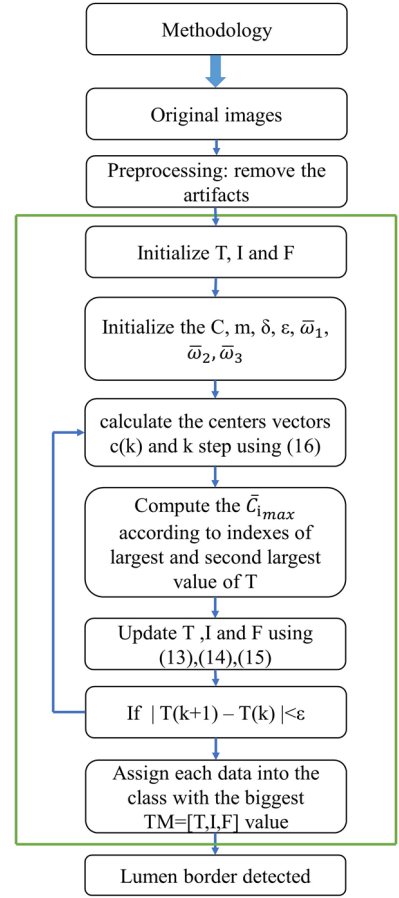

Fig.2. shows the steps of algorithm

*B. Evaluation metrics*

To study the effect of implementing the proposed approach by using different evaluation measures, we devise an experiment in which we compare the final accuracy of our method while using the following various measures: AD, HD, JACC, PAD, and DI.

AD is defined as the average distance between automatically detected lumen and the manual lumen boundary [24].

$$AD = 1 - \left(\frac{A_{automatic} \cap A_{manual}}{(A_{automatic} + A_{manual}) - (A_{automatic} \cap A_{manual})}\right) \quad (18)$$

where $A_{automatic}$ is the boundaries which have been detected by the proposed method and $A_{manual}$ is the borders which have been defined by experts.

HD which is the maximum distance between the detected lumen and the manual lumen boundary [25].

$$HD(A_{automatic}, A_{manual}) = \max\{\max[d(a,b)]\}$$
$$a \in A_{automatic}, b \in A_{manual} \quad (19)$$

In equation (19), a and b are curve points of $A_{automatic}$ and $A_{manual}$, respectively, and d(a, b) is the Euclidean distance.

The overlapping area ratio between $A_{automatic}$ and $A_{manual}$ can be defined by JACC. JACC value is between 0 and 1 for the worst and the best situation, respectively [25].

$$JACC = \frac{A_{automatic} \cap A_{manual}}{A_{automatic} \cup A_{manual}} \quad (20)$$

PAD calculates the difference between $A_{automatic}$ and $A_{manual}$ [26].

$$PAD = \frac{|A_{automatic} - A_{manual}|}{A_{manual}} \quad (21)$$

The similarity between the region $A_{automatic}$ and $A_{manual}$ is indicated by the Dice similarity index (DI). The Dice Similarity Index is as follows:

$$DI(A, B) = \frac{2|A_{automatic} \cap A_{manual}|}{|A_{automatic}| + |A_{manual}|} \quad (22)$$

Where, $|A_{automatic}|$ and $|A_{manual}|$ illustrate the number of pixels of the lumen area in image $A_{automatic}$ and $A_{manual}$. $|A_{automatic} \cap A_{manual}|$ shows the number of pixels of the overlapping lumen area of image $A_{automatic}$ and $A_{manual}$. DI value is between 0 and 1 for the worst and the best situation, respectively [27].

## IV. EXPERIMENTAL RESULTS

### A. Dataset

The dataset includes 138 frames from a single patient. An imaging system with enabled optical coherence tomography and NIRF technologies was used to obtain imaging from a single pullback scanning, by employing a custom-made catheter. The diameter of the catheter is 0.87 mm. High-resolution cross-sectional images at an A-line rate of 120 kHz were provided by an optical coherence tomography system which its center has 1290 nm wavelength [28].

### B. Figures and Tables

The NCM clustering method was applied over IVOCT images. The method labels the image pixels into multiple cluster regions. The dataset is collected for a single patient and includes a set of 138 IVOCT images. The image clustering was performed on a computer with an Intel Core i3 as CPU, 4 GB of RAM, Windows 7 64bits as the operating system and MATLAB (2018a) software. After performing the NCM method over the given dataset, the average value for DI, JACC, PAD, AD and HD evaluation measures, which are previously defined, are 0.985, 0.972, 0.019, 0.076 mm and 0.32 mm respectively. As it is depicted by the following figures, our approach achieves a higher accuracy rate than the previous works.

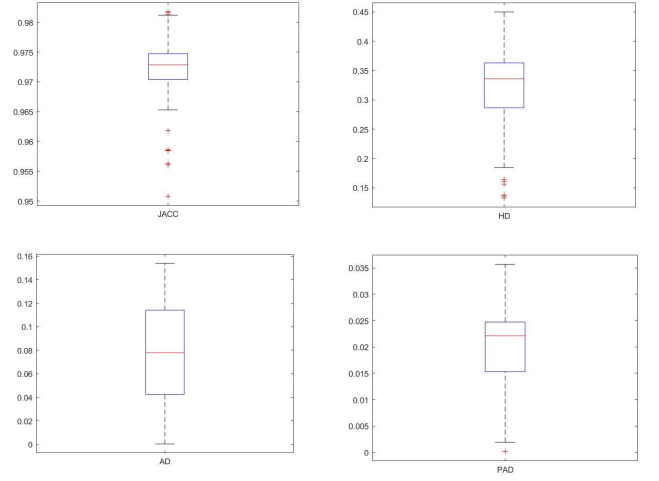

Fig.3. shows the statistical indexes of evaluation metrics.

The following figure represents the final results of our method from another perspective. In this figure, the first, second and third columns sketch the original input image, the manually detected lumen and the automatic lumen detection by the NCM method respectively. As it is clearly obvious, the lumen is detected with high accuracy (the red circular shape).

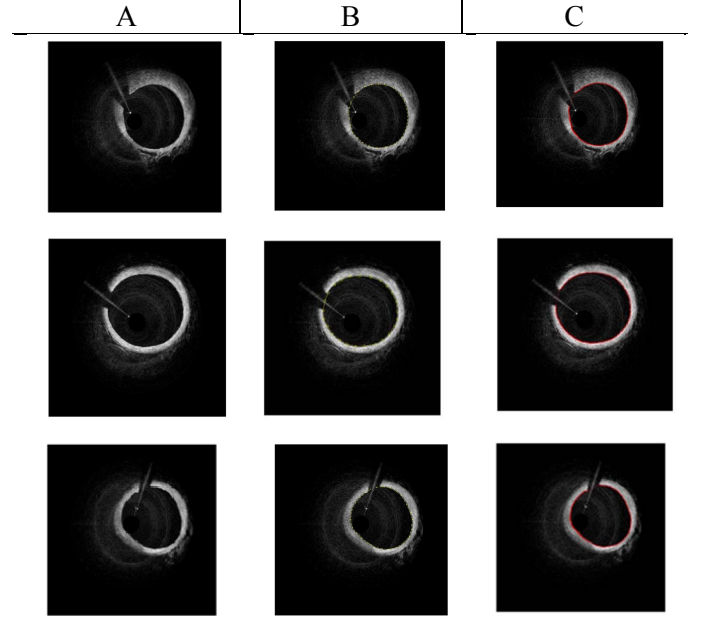

Fig.4. Three IVOCT frame (column A), along with the manually detected lumen (column B) and automatically detected lumen boundaries by NCM method (column C).

The JACC, PAD, DI, HD, and AD evaluation metrics' mean values of proposed method are shown in table I:

TABLE I. METHODS COMPARISON

| | JACC % | HD (mm) | AD (mm$^2$) | PAD | DI % |
|---|---|---|---|---|---|
| binary morphological reconstruction [29] | 95.6 | --- | --- | --- | 97.8±2.16 |
| Convolutional Neural Network [7]: Dataset 1 Dataset 2 Dataset 3 | 90.80 93.83 97.6±2.3 | --- --- --- | --- --- --- | --- --- --- | 94.34 96.73 98.8 ± 1.2 |
| Wavelet and Mathematical Morphology [30] | --- | 0.084 | 0.06 | --- | --- |
| Proposed method | 97.2 ± 2.1 | 0.32 ± 0.18 | 0.076 ±0.077 | 0.019 ±0.01 | 98.5 ± 1.36 |

The evaluation measures, for the previously defined dataset, are evaluated by our Automatic lumen detection using NCM, compared with binary morphological reconstruction [29], Convolutional neural network [7], and Wavelet and mathematical morphology [30]. The final results have been shown in Table I. It is noteworthy that all the values in the table are computed by averaging the methods' result for the whole dataset instances (138 instances for a single patient).

IVOCT papers have not used PAD metric in their evaluations yet, although this metric is essential for medical analysis. As shown in Table I, the mean value of PAD is 0.019 employing our proposed method.

DI and JACC metric achieved decent percentages, and PAD as well as AD obtained small and proper quantity. However, the guide wire effect caused high-intensity points in the lumen, resulting in inappropriate value for HD.

## V. CONCLUSION

In this study, we used one of the precise clustering methods in the NS domain, whereby the boundary and noise can be identified using the NCM cost function. The purpose of this clustering was to find the lumen boundaries in the IVOCT images, which, due to the noise removal and the use of the uncertainty, have been carefully designed to detect these boundaries.

In this work, the accuracy of FCM is improved. In our future works, NCM can be combined with deep learning networks such as the convolutional neural network for image clustering. Moreover, a new method can be presented in order to define the number of clusters automatically.

REFERENCES

[1] J. Perk, G. De Backer, and H. Gohlke, "European Guidelines on cardiovascular disease prevention in clinical practice (version 2012)The Fifth Joint Task Force of the European Society of Cardiology and Other Societies on Cardiovascular Disease Prevention in Clinical Practice (constituted by representatives of nine societies and by invited experts)Developed with the special contribution of the European Association for Cardiovascular Prevention & Rehabilitation (EACPR)†," *European Heart Journal* vol. 33, pp. 1635-1701, 2012.

[2] Y. Cao, K. Cheng, X. Qin, Q. Yin, J. Li, R. Zhu, *et al.*, "Automatic Lumen Segmentation in Intravascular Optical Coherence Tomography Images Using Level Set," *Computational and Mathematical Methods in Medicine,* vol. 2017, p. 11, 2017.

[3] J. G. Fujimoto, M. E. Brezinski, G. J. Tearney, S. A. Boppart, B. Bouma, M. R. Hee, *et al.*, "Optical biopsy and imaging using optical coherence tomography," *Nature Medicine,* vol. 1, p. 970, 09/01/online 1995.

[4] F. Dubuisson, C. Kauffmann, P. Motreff, and L. Sarry, "In Vivo OCT Coronary Imaging Augmented with Stent Reendothelialization Score," in *Medical Image Computing and Computer-Assisted Intervention – MICCAI 2009*, Berlin, Heidelberg, 2009, pp. 475-482.

[5] A. Wang and X. Tang, *Automatic segmentation of vessel lumen in intravascular optical coherence tomography images*, 2016.

[6] H. Azarnoush, S. Vergnole, B. Boulet, R. DiRaddo, and G. Lamouche, "Real-Time Control of Angioplasty Balloon Inflation Based on Feedback From Intravascular Optical Coherence Tomography: Preliminary Study on an Artery Phantom," *IEEE Transactions on Biomedical Engineering,* vol. 59, pp. 697-705, 2012.

[7] M. Miyagawa, M. G. F. Costa, M. A. Gutierrez, J. P. G. F. Costa, and C. F. F. C. Filho, "Lumen Segmentation in Optical Coherence Tomography Images using Convolutional Neural Network," in *2018 40th Annual International Conference of the IEEE Engineering in Medicine and Biology Society (EMBC)*, 2018, pp. 600-603.

[8] A. G. Roy, S. Conjeti, S. G. Carlier, P. K. Dutta, A. Kastrati, A. F. Laine, *et al.*, "Lumen Segmentation in Intravascular Optical Coherence Tomography Using Backscattering Tracked and Initialized Random

Walks," *IEEE Journal of Biomedical and Health Informatics,* vol. 20, pp. 606-614, 2016.

[9] H. Modanloujouybari, A. Ayatollahi, and A. Kermani, "Vessel wall detection in the images of intravascular Optical coherence tomography based on the graph cut segmentation," in *2017 Iranian Conference on Electrical Engineering (ICEE)*, 2017, pp. 39-44.

[10] S. Su, Z. Hu, Q. Lin, W. K. Hau, Z. Gao, and H. Zhang, "An artificial neural network method for lumen and media-adventitia border detection in IVUS," *Computerized Medical Imaging and Graphics,* vol. 57, pp. 29-39, 2017/04/01/ 2017.

[11] E. Rashno, A. Akbari, and B. Nasersharif, "A Convolutional Neural Network model based on Neutrosophy for Noisy Speech Recognition," *arXiv preprint arXiv:1901.10629,* 2019.

[12] E. Dos Santos, M. Yoshizawa, A. Tanaka, Y. Saijo, and T. Iwamoto, *Detection of Luminal Contour Using Fuzzy Clustering and Mathematical Morphology in Intravascular Ultrasound Images* vol. 4, 2005.

[13] E. Rashno, B. Minaei-Bidgolia, and Y. Guo, "An effective clustering method based on data indeterminacy in neutrosophic set domain," *arXiv preprint arXiv:1812.11034,* 2018.

[14] Z. Wu and R. Leahy, "An optimal graph theoretic approach to data clustering: theory and its application to image segmentation," *IEEE Transactions on Pattern Analysis and Machine Intelligence,* vol. 15, pp. 1101-1113, 1993.

[15] A. K. Jain, "Data clustering: 50 years beyond K-means," *Pattern Recognition Letters,* vol. 31, pp. 651-666, 2010/06/01/ 2010.

[16] Y. Guo and A. Sengur, *NCM: Neutrosophic c-means clustering algorithm*, 2015.

[17] E. Rashno, S. S. Norouzi, B. Minaei-bidgoli, and Y. Guo, "Certainty of outlier and boundary points processing in data mining," *arXiv preprint arXiv:1812.11045,* 2018.

[18] Y. Guo and H. D. Cheng, *New neutrosophic approach to image segmentation* vol. 42, 2009.

[19] M. A. Jaffar, N. Naveed, B. Ahmed, A. Hussain, and A. M. Mirza, "Fuzzy c-means clustering with spatial information for color image segmentation," in *2009 Third International Conference on Electrical Engineering*, 2009, pp. 1-6.

[20] K.-S. Chuang, H.-L. Tzeng, S. Chen, J. Wu, and T.-J. Chen, "Fuzzy c-means clustering with spatial information for image segmentation," *Computerized medical imaging and graphics : the official journal of the Computerized Medical Imaging Society,* vol. 30, pp. 9-15, 2006/01// 2006.

[21] K. K. Chintalapudi and M. Kam, "A noise-resistant fuzzy c means algorithm for clustering," in *1998 IEEE International Conference on Fuzzy Systems Proceedings. IEEE World Congress on Computational Intelligence (Cat. No.98CH36228)*, 1998, pp. 1458-1463 vol.2.

[22] H. D. Cheng and Yanhuiguo, *A NEW NEUTROSOPHIC APPROACH TO IMAGE THRESHOLDING* vol. 04, 2011.

[23] A. Rashno, D. D. Koozekanani, P. M. Drayna, B. Nazari, S. Sadri, H. Rabbani*, et al.*, "Fully Automated Segmentation of Fluid/Cyst Regions in Optical Coherence Tomography Images With Diabetic Macular Edema Using Neutrosophic Sets and Graph Algorithms," *IEEE Transactions on Biomedical Engineering,* vol. 65, pp. 989-1001, 2018.

[24] H. Sofian, S. Muhammad, J. Than, C. Ming, and N. Noor, *Lumen Coronary Artery Border Detection Using Texture and Chi-square Classification*, 2015.

[25] S. Balocco, C. Gatta, F. Ciompi, A. Wahle, P. Radeva, S. Carlier*, et al.*, "Standardized evaluation methodology and reference database for evaluating IVUS image segmentation," *Computerized Medical Imaging and Graphics,* vol. 38, pp. 70-90, 2014/03/01/ 2014.

[26] A. Kermani and A. Ayatollahi, "A new nonparametric statistical approach to detect lumen and Media-Adventitia borders in intravascular ultrasound frames," *Computers in Biology and Medicine,* vol. 104, pp. 10-28, 2019/01/01/ 2019.

[27] M. Xu, J. Cheng, D. W. K. Wong, J. Liu, A. Taruya, and A. Tanaka, "Graph based lumen segmentation in optical coherence tomography images," in *2015 10th International Conference on Information, Communications and Signal Processing (ICICS)*, 2015, pp. 1-5.

[28] S. Lee, W. Lee Min, S. Cho Han, W. Song Joon, S. Nam Hyeong, J. Oh Dong*, et al.*, "Fully Integrated High-Speed Intravascular Optical Coherence Tomography/Near-Infrared Fluorescence Structural/Molecular Imaging In Vivo Using a Clinically Available Near-Infrared Fluorescence–Emitting Indocyanine Green to Detect Inflamed Lipid-Rich Atheromata in Coronary-Sized Vessels," *Circulation: Cardiovascular Interventions,* vol. 7, pp. 560-569, 2014/08/01 2014.

[29] M. C. Moraes, D. A. C. Cardenas, and S. S. Furuie, "Automatic lumen segmentation in IVOCT images using binary morphological reconstruction," *BioMedical Engineering OnLine,* vol. 12, p. 78, 2013/08/09 2013.

[30] M. C. Moraes, D. A. C. Cardenas, and S. S. Furuie, "Automatic IOCT lumen segmentation using Wavelet and Mathematical Morphology," in *2012 Computing in Cardiology*, 2012, pp. 545-548.